\documentstyle[11pt]{article}

  \def\pp{{\mathchoice
          {
              \kern 1pt%
              \raise 1pt
              \vbox{\hrule width5pt height0.4pt depth0pt
                    \kern -2pt
                    \hbox{\kern 2.3pt
                          \vrule width0.4pt height6pt depth0pt
                          }
                    \kern -2pt
                    \hrule width5pt height0.4pt depth0pt}%
                    \kern 1pt
           }
            {
              \kern 1pt%
              \raise 1pt
              \vbox{\hrule width4.3pt height0.4pt depth0pt
                    \kern -1.8pt
                    \hbox{\kern 1.95pt
                          \vrule width0.4pt height5.4pt depth0pt
                          }
                    \kern -1.8pt
                    \hrule width4.3pt height0.4pt depth0pt}%
                    \kern 1pt
            }
            {
              \kern 0.5pt%
              \raise 1pt
              \vbox{\hrule width4.0pt height0.3pt depth0pt
                    \kern -1.9pt  
                    \hbox{\kern 1.85pt
                          \vrule width0.3pt height5.7pt depth0pt
                          }
                    \kern -1.9pt
                    \hrule width4.0pt height0.3pt depth0pt}%
                    \kern 0.5pt
            }
            {
              \kern 0.5pt%
              \raise 1pt
              \vbox{\hrule width3.6pt height0.3pt depth0pt
                    \kern -1.5pt
                    \hbox{\kern 1.65pt
                          \vrule width0.3pt height4.5pt depth0pt
                          }
                    \kern -1.5pt
                    \hrule width3.6pt height0.3pt depth0pt}%
                    \kern 0.5pt
            }
        }}

  \def\mm{{\mathchoice
                       {
                             \kern 1pt
               \raise 1pt    \vbox{\hrule width5pt height0.4pt depth0pt
                                  \kern 2pt
                                  \hrule width5pt height0.4pt depth0pt}
                             \kern 1pt}
                       {
                            \kern 1pt
               \raise 1pt \vbox{\hrule width4.3pt height0.4pt depth0pt
                                  \kern 1.8pt
                                  \hrule width4.3pt height0.4pt depth0pt}
                             \kern 1pt}
                       {
                            \kern 0.5pt
               \raise 1pt
                            \vbox{\hrule width4.0pt height0.3pt depth0pt
                                  \kern 1.9pt
                                  \hrule width4.0pt height0.3pt depth0pt}
                            \kern 1pt}
                       {
                           \kern 0.5pt
             \raise 1pt  \vbox{\hrule width3.6pt height0.3pt depth0pt
                                  \kern 1.5pt
                                  \hrule width3.6pt height0.3pt depth0pt}
                           \kern 0.5pt}
                       }}

\catcode`@=11
\def\un#1{\relax\ifmmode\@@underline#1\else
        $\@@underline{\hbox{#1}}$\relax\fi}
\catcode`@=12

\let\du=\du                     

\def\a{\alpha}
\def\b{\beta}

\def\d{\delta}

\def\f{\phi}
\def\g{\gamma}
\def\h{\eta}

\def\j{\psi}
\def\k{\kappa}
\def\l{\lambda}
\def\m{\mu}
\def\n{\nu}
\def\o{\omega}
\def\p{\pi}
\def\q{\theta}

\def\x{\xi}

\def\O{\Omega}

\def\ve{\varepsilon}
\def\vf{\varphi}

\def\ck{{\cal K}}

\def\cm{{\cal M}}

\def\bo{{\raise-.5ex\hbox{\large$\Box$}}}               
\def\pa{\partial}                                       
\def\TH{{\raise.2ex\hbox{$\displaystyle \bigodot$}\mskip-4.7mu \llap H \;}}
\def\face{{\raise.2ex\hbox{$\displaystyle \bigodot$}\mskip-2.2mu \llap {$\ddot
        \smile$}}}                                      

\def\sp#1{{}^{#1}}                              
\def\abs#1{\left| #1\right|}                    
\def\leftrightarrowfill{$\mathsurround=0pt \mathord\leftarrow \mkern-6mu
        \cleaders\hbox{$\mkern-2mu \mathord- \mkern-2mu$}\hfill
        \mkern-6mu \mathord\rightarrow$}
\def\dvec#1{\vbox{\ialign{##\crcr
        \leftrightarrowfill\crcr\noalign{\kern-1pt\nointerlineskip}
        $\hfil\displaystyle{#1}\hfil$\crcr}}}           

\def\frac#1#2{{\textstyle{#1\over\vphantom2\smash{\raise.20ex
        \hbox{$\scriptstyle{#2}$}}}}}                   
\def\sfrac#1#2{{\vphantom1\smash{\lower.5ex\hbox{\small$#1$}}\over
        \vphantom1\smash{\raise.4ex\hbox{\small$#2$}}}} 
\def\bfrac#1#2{{\vphantom1\smash{\lower.5ex\hbox{$#1$}}\over
        \vphantom1\smash{\raise.3ex\hbox{$#2$}}}}       
\def\afrac#1#2{{\vphantom1\smash{\lower.5ex\hbox{$#1$}}\over#2}}    

\def\[{\lfloor{\hskip 0.35pt}\!\!\!\lceil}
\def\]{\rfloor{\hskip 0.35pt}\!\!\!\rceil}

\def\du#1#2{_{#1}{}^{#2}}

\def\Tr{{\rm Tr}}

\def\un{\underline}
\def\fracmm#1#2{{{#1}\over{#2}}}

\def\low#1{{\raise -3pt\hbox{${\hskip 0.75pt}\!_{#1}$}}}

\newskip\humongous \humongous=0pt plus 1000pt minus 1000pt
\def\caja{\mathsurround=0pt}
\def\eqalign#1{\,\vcenter{\openup2\jot \caja
        \ialign{\strut \hfil$\displaystyle{##}$&$
        \displaystyle{{}##}$\hfil\crcr#1\crcr}}\,}
\newif\ifdtup

\def\ref#1{$\sp{#1)}$}

\parskip=\medskipamount                 
\thicklines                         

\date{}
\title{Do the critical (2,2) strings know about \\
        a supergravity in 2+10 dimensions~?}
\author{S. V. Ketov${}^1$\\
${}^1${\it Institut f\"ur Theoretische Physik, Universit\"at Hannover, 
Germany}}

\begin{document}

\maketitle

\begin{center}
Talk given at the XIIth International Congress of Mathematical Physics,\\
July 13--19, 1997, in Brisbane, Australia, and the Second International \\
Conference on Quantum Field Theory and Gravity, July 29--August 2, \\
1997, in Tomsk, Russia (=Hannover preprint ITP-UH-27/97, October'97) 
\end{center}
\begin{abstract}
{\bf 
The effective field equations of motion for a mixed theory of open and closed 
(2,2) world-sheet supersymmetric critical strings are shown to be integrable 
in the case of an {\it abelian} gauge group. The Born-Infeld-type effective 
action in 2+2 dimensions is intrinsically non-covariant, and it can be
interpreted as (a part of) the F-brane world-volume action. The covariant
F-brane action is unambiguously restored by its maximal (N=8) world-volume 
supersymmetry. The 32 supercharges, the local $SO(2,1)\otimes SO(8)$ and the
rigid $SL(2,R)$ symmetries of the F-brane action naturally suggest its 
interpretation as the hypothetical (non-covariant) self-dual `heterotic' (1,0)
supergravity in 2+10 dimensions.}
\end{abstract}

The {\it closed} (2,2) world-sheet supersymmetric critical string theory is 
known to have the interpretation of being a theory of {\it self-dual gravity}
(SDG)~\cite{ov}. Similarly, the {\it open} (2,2) critical string theory can be
interpreted as a {\it self-dual Yang-Mills} (SDYM) theory~\cite{mar}. Since 
open strings can `create' closed strings which, in their turn, can interact 
with the open strings, there are quantum corrections to the effective field 
equations of the open (2,2) string theory. Because of the  `topological' 
nature of the (2,2) string theories, {\it only} 3-point tree string amplitudes 
are non-vanishing and local. As a result, quantum perturbative corrections in 
the mixed theory of open {\it and} closed (2,2) strings are  under control. In
particular, the SDYM equations receive corrections from diagrams with internal
gravitons, so that they become the YM self-duality equations on a K\"ahler
background~\cite{mar}. In particular, they respect integrability. Contrary
to the SDYM equations and naive expectations, the effective gravitational
equations of motion in the mixed (open/closed) (2,2) string theory get modified
 so that the resulting `spacetime' is no longer self-dual~\cite{mar,svk}. The 
integrability is nevertheless maintained in the case of an {\it abelian} gauge
 group~\cite{svk}. The effective field theory is different from the standard
Einstein-Maxwell system describing the interaction of a non-linear graviton 
with a photon field. Instead, it is of the Born-Infeld-type, i.e. non-linear 
with respect to the {\it both} fields. 

The (2,2) strings are strings with {\it two} world-sheet supersymmetries, both 
for the left- and right-moving degrees of freedom.~\footnote{(2,1) and (2,0) 
heterotic strings can also be defined \cite{ovh}. Since the heterotic strings 
have to live in a \newline ${~~~~~}$ (2+1)- or (1+1)-dimensional target space,
where self-duality is lost (or hidden, at least), we do not \newline ${~~~~~}$
consider them here (see, however, ref.~\cite{klm}).} The critical open and 
closed (2,2) strings live in four real dimensions, with the signature $(2,2)$.
 Their physical spectrum consists of a single massless particle, which can be 
assigned in the adjoint of a gauge group $G$ in the open string case.

The only non-vanishing $(2,2)$ string tree scattering amplitudes are 3-point
trees, while all higher $n$-point functions should vanish due to kinematical 
reasons in 2+2 dimensions. Tree-level calculations of string amplitudes do not
require a heavy mashinery of the BRST quantization~\cite{hann}, or the (N=4)
topological methods~\cite{bv}. The vertex operator for a (2,2) closed string 
particle of momentum $k$ simply reads in the (2,2) world-sheet superspace as
$$ V_{\rm c} =\fracmm{\k}{\p}\exp\left\{ i\left(k\cdot\bar{Z}+\bar{k}\cdot Z
\right)\right\}~,\eqno(1)$$
where $\k$ is the $(2,2)$ closed string coupling constant, and 
$Z^i(x,\bar{x},\q,\bar{\q})$ are complex (2,2) chiral superfields.~\footnote{
Throughout the paper, complex coordinates $(x,\bar{x})$ are used for the string
world-sheet, while $(z^i,\bar{z}^{\bar{i}})$ \newline ${~~~~~}$ denote complex 
coordinates of the (2,2) string target space, $i=1,2$.}

When using the $(2,2)$ super-M\"obius invariance of the $(2,2)$ super-Riemann
sphere, it is not difficult to calculate the correlation function of three 
$V_{\rm c}\,$. One finds \cite{ov}
$$ A_{\rm ccc}= \k c_{23}^2~,\quad {\rm where}\quad
c_{23}\equiv \left(k_2\cdot\bar{k}_3-\bar{k}_2\cdot k_3\right)~.\eqno(2)$$
One can check that the $A_{\rm ccc}$ is totally (crossing) symmetric on-shell,
but it is only invariant under the subgroup $U(1,1)\cong SL(2,{\bf R})\otimes 
U(1)$ of the full `Lorentz' group $SO(2,2)\cong SL(2,{\bf R})\otimes 
SL(2,{\bf R})'\cong SO(1,2)\otimes SO(1,2)'$ in 2+2 dimensions.

Since all higher correlators vanish, the local 3-point function (2) alone 
determines the (perturbatively) {\it exact} effective action \cite{ov},
$$ S_{\rm P} =\int d^{2+2}z\,\left( \fracmm{1}{2}\h^{i\bar{j}}\pa_i\f
\bar{\pa}_{\bar{j}}\f + \fracmm{2\k}{3}\f\pa\bar{\pa}\f\wedge 
\pa\bar{\pa}\f\right)~,\eqno(3)$$
which is known as the {\it Pleba\'nski} action for self-dual gravity (SDG).  
Hence, the massless `scalar' of the closed string theory can be identified 
with a deformation of the K\"ahler potential $K$ of the self-dual 
(=K\"ahler + Ricci-flat) gravity~\cite{ov}, where
$$ K=\h_{i\bar{j}}z^i\bar{z}^{\bar{j}}+4\k\f~,\qquad 
\h_{i\bar{j}}=\h^{i\bar{j}}=
\left( \begin{array}{cc} 1 & 0 \\ 0 & -1 \end{array}\right)~.
\eqno(4)$$
The (2,2) closed string target space metric is therefore given by 
$$ g_{i\bar{j}}=\pa_i\bar{\pa}_{\bar{j}}K=\h_{i\bar{j}}
+4\k \pa_i\bar{\pa}_{\bar{j}}\f~.\eqno(5)$$

Similarly, in the open (2,2) string case, when using the 
$N=(2,2)$ superspace vertex
$$ V_{\rm o} =g\exp\left\{ i\left(k\cdot\bar{Z}+\bar{k}\cdot Z
\right)\right\}~,\eqno(6)$$
assigned to the boundary of the $(2,2)$ supersymmetric upper-half-plane 
(or $(2,2)$ super-disc) with proper boundary conditions, one finds the 
three-point function~\cite{mar}
$$ A_{\rm ooo} =-igc_{23}f^{abc}~,\eqno(7)$$
which is essentially a `square root' of $A_{\rm ccc}$, while  $f^{abc}$
are structure constants of $G$. The $ A_{\rm ooo}$ can be obtained from the 
effective action~\cite{mar}
$$ S_{\rm DNS} = \int d^{2+2}z\,\h^{i\bar{j}}\left( \fracmm{1}{2}\pa_i\vf^a
\bar{\pa}_{\bar{j}}\vf^a 
- i\fracmm{g}{3}f^{abc}\vf^a\pa_i\vf^b\bar{\pa}_{\bar{j}}
\vf^c\right)+\,~\ldots \eqno(8)$$
Requiring all the higher-point amplitudes to vanish in the {\it field} theory 
(8) determines the additional local $n$-point interactions, $n>3$, which were
denoted by dots in eq.~(8). The full action $S_{\rm DNS}$ appears to be the 
{\it Donaldson-Nair-Schiff} (DNS) action~\cite{dns}. The DNS equation of motion
 is known as the {\it Yang} equation~\cite{yang}~:
$$ \h^{i\bar{j}} \bar{\pa}_{\bar{j}}\left( e^{-2ig\vf}\pa_i e^{2ig\vf}\right)
=0~,\eqno(9)$$
where the matrix $\vf$ is Lie algebra-valued, $\vf=\vf^at^a$, and the Lie 
algebra generators $t^a$ of $G$ are taken to be anti-hermitian. The DNS action
is known to be dual (in the field theory sense) to the {\it Leznov}  action
\cite{lez}, which has only {\it cubic} interaction, and whose equation of 
motion also describes the SDYM.~\footnote{The Leznov action is the effective 
field theory action of open (2,2) strings when the {\it world-sheet instanton} 
\newline ${~~~~~}$ corrections are included~\cite{slecht}. All the higher 
point functions vanish in the Leznov quantum field theory~\cite{parkes}.} 

When open strings join together, they form closed strings. In particular, the 
coupling constants of the closed and open strings are related,
$$ \k \sim \sqrt{\hbar}\,g^2~.\eqno(10)$$
The mixed (2,2) string tree amplitudes were also calculated ~\cite{mar}. The
only non-vanishing 3-point mixed amplitude is given by
$$ A_{\rm ooc} =\fracmm{\k}{\p} \d^{ab}c_{23}^2 \int^{+\infty}_{-\infty}dx\,
\fracmm{1}{x^2+1}=\k\d^{ab}c_{23}^2~,\eqno(11)$$
where the integration over the position $x$ of one of the open string vertices 
goes along the border of the upper-half-plane (= real line). All higher 
$n$-point mixed amplitudes, $n\geq 4$, are believed to vanish too. The extra
term in the open string effective action, that reproduces $A_{\rm ooc}$, reads 
~\cite{mar}:
$$ S_{\rm mixed} = \int d^{2+2}z\,\left( 2\k\f\pa\bar{\pa}\vf^a\wedge 
\pa\bar{\pa}\vf^a\right)~.\eqno(12)$$
The complete non-abelian effective action can be determined by demanding all 
higher-point amplitudes to vanish in the {\it field} theory describing the 
mixed (2,2) strings, order by order in $n$. Rescaling $\f$ by a factor of 
$4\k$, and $\vf$ by a factor of $g$, one finds the effective equations of
motion in the form~\cite{mar,svk}:
$$g^{i\bar{j}}(\f)\bar{\pa}_{\bar{j}}\left( e^{-2i\vf}\pa_i e^{2i\vf}\right)
=0~,\eqno(13)$$
and
$$-\det g_{i\bar{j}}=+1 +\fracmm{2\k^2}{g^2}\Tr\left(F_{i\bar{j}}F^{i\bar{j}}
\right)~,\eqno(14)$$
where $F_{i\bar{j}}$ is the YM field strength of the YM gauge fields
$$A\equiv e^{-i\vf}\pa e^{i\vf}~, \qquad \bar{A}\equiv e^{i\vf}\bar{\pa}
 e^{-i\vf}~, \eqno(15)$$
$g_{i\bar{j}}=\h_{i\bar{j}}+\pa_i\bar{\pa}_{\bar{j}}\f$ is a K\"ahler metric,
$g^{i\bar{j}}$ is its inverse, and the indices $(i,\bar{j})$ are raised and 
lowered by using the totally antisymmetric Levi-Civita symbols $\ve^{ij}$,  
$\ve^{\bar{i}\bar{j}}$, and $\ve_{ij}$, $\ve_{\bar{i}\bar{j}}$
$(\ve_{12}=\ve^{12}=1)$.

Eq.~(13) is just the Yang equation describing the SDYM on a curved K\"ahler 
gravitational background. Associated with the K\"ahler metric
$$ ds^2= 2g_{i\bar{j}}dz^id\bar{z}^{\bar{j}}\equiv 2K_{,i\bar{j}}dz^i
d\bar{z}^{\bar{j}}~,\eqno(16)$$
is the fundamental (K\"ahler) closed two-form
$$ \O=g_{i\bar{j}}dz^i\wedge d\bar{z}^{\bar{j}}\equiv K_{,i\bar{j}}dz^i\wedge
d\bar{z}^{\bar{j}}~,\eqno(17)$$
where $K$ is the (locally defined) K\"ahler potential, and all subscripts 
after a comma denote partial differentiations. We regard the complex 
coordinates $(z^i,\bar{z}^{\bar{i}})$ as independent variables, so that our 
complexified  `spacetime' $\cm$ is locally a direct product of two 
2-dimensional complex manifolds $\cm\cong M_2\otimes\bar{M}_2$, where both 
 $M_2$ and $\bar{M}_2$ are endowed with complex structures, i.e. possess
 closed non-degenerate two-forms $\o$ and $\bar{\o}$, respectively.~\footnote{
The normalization of the holomorphic two-forms $\o$ and $\bar{\o}$ is fixed by
 the flat `spacetime' limit where one \newline
${~~~~~}$ has $\o=dz^1\wedge dz^2$ and $\bar{\o}=d\bar{z}^{\bar{1}}\wedge 
d\bar{z}^{\bar{2}}$.} Hence, the effective equations of motion (13) and (14) 
of the mixed (2,2) string theory can be rewritten to the coordinate-independent
form~\cite{mar}:
$$ \O\wedge F=0~,\qquad
 \O\wedge \O +\fracmm{4\k^2}{g^2}\Tr (F\wedge F)=2\o\wedge\bar{\o}~,
\eqno(18)$$
where $F$ is the YM Lie algebra-valued field strength two-form satisfying
$$\o\wedge F=\bar{\o}\wedge F=0~.\eqno(19)$$

The first equation (18) and eq.~(19) are just the {\it self-dual} Yang-Mills 
equations in a K\"ahler `spacetime'. They are well-known to be integrable, 
while their solutions describe Yang-Mills instantons. In particular, one
can always locally change the flat SDYM equations of motion into the SDYM 
equations on a curved K\"ahler background by a diffeomorphism transformation
compatible with the K\"ahler structure.

The integrability condition for the gravitational equations of motion in the
complexified `spacetime' is known to be precisely equivalent to the 
(anti)self-duality of the {\it Weyl} curvature tensor~\cite{pen}. The famous
twistor construction of Penrose~\cite{pen} transforms the problem of solving
the non-linear partial differential equations of conformally self-dual gravity
into the standard Riemann-Hilbert problem of patching together certain
holomorphic data.

As far as the {\it K\"ahler} spaces are concerned, the self-duality of the Weyl
tensor is precisely equivalent to the vanishing Ricci {\it scalar} curvature
\cite{fla,bpl}, while the Ricci tensor itself is simply related to the
  K\"ahler metric as
$$ R_{i\bar{j}}=\pa_i\bar{\pa}_{\bar{j}}\log\det(g_{k\bar{k}})~.\eqno(20)$$
One easily finds that
$$  R_{i\bar{j}}=\pa_i\bar{\pa}_{\bar{j}}\log\left[ 1+\fracmm{2\k^2}{g^2}\Tr
(F_{i\bar{j}}F^{i\bar{j}})\right]~,\eqno(21)$$
and, therefore, it is quite obvious that the `matter' stress-energy tensor, 
that has to be equal to the Einstein tensor to be constructed out of eqs.~(20)
 and (21), does {\it not} vanish.  It is to be compared to the  standard
coupled Eistein-Yang-Mills system, where the YM  stress-energy tensor
is quadratic with respect to the YM field strength, and it vanishes under the
SDYM condition. In our case, the YM stress-energy tensor is not even polynomial
in the YM field strength, and it has to correspond to a non-polynomial (in $F$)
effective action.

In order to understand the meaning of eq.~(14) or the second eq.~(18), let's 
rewrite  it to the form:
$$ \det(g_{i\bar{j}})+\fracmm{2\k^2}{g^2}\Tr\det(F_{i\bar{j}})=-1~,\eqno(22)$$
where both determinants are two-dimensional. Given an {\it abelian} field 
strength $F$ satisfying the {\it self-duality} condition  or, equivalently,
$g_{1\bar{1}}F_{2\bar{2}}+g_{2\bar{2}}F_{1\bar{1}}-g_{1\bar{2}}F_{2\bar{1}}-
g_{2\bar{1}}F_{1\bar{2}}=0$, there is a remarkable identity~\cite{svk}
$$ \det(g) + \fracmm{2\k^2}{g^2}\det(F) = 
\det\left(g+\fracmm{\k\sqrt{2}}{g}F\right)~.\eqno(23)$$
In addition, eq.~(19) in the abelian case implies
$A=i\pa\vf$, $\bar{A}=-i\bar{\pa}\vf$, and, hence, $F=2i\pa\bar{\pa}\vf$. We
are now in a position to represent eq.~(22) as the Pleba\'nski heavenly
equation
$$ \det\left(\pa\bar{\pa}\ck\right)=-1~,\eqno(24)$$
in terms of the formal {\it complex} K\"ahler potential
$$\ck\equiv K + i\fracmm{2\sqrt{2}\k}{g}\vf~,\eqno(25)$$
whose imaginary part is a harmonic function (because of the self-duality of 
$F$), and of the order $\hbar^{1/2}g$. Eq.~(25) is the consistency condition 
of the linear system
$$\eqalign{
L_{\bar{1}}\j\equiv \left[\bar{\pa}_{\bar{1}}+i\l\bar{B}_{\bar{1}}\right]\j
\equiv 
&  \left[\bar{\pa}_{\bar{1}}+i\l\left(\ck_{,2\bar{1}}\pa_1-\ck_{,1\bar{1}}\pa_2
\right)\right]\j=0~,\cr
L_{\bar{2}}\j\equiv \left[\bar{\pa}_{\bar{2}}+i\l\bar{B}_{\bar{2}}\right]\j
\equiv
&  \left[\bar{\pa}_{\bar{2}}+i\l\left(\ck_{,2\bar{2}}\pa_1-\ck_{,1\bar{2}}\pa_2
\right)\right]\j=0~,\cr}\eqno(26)$$
where $\l$ is a complex spectral parameter. Eq.~(26) amounts to the 
 Frobenius integrability, just like that in the usual 
case of the Pleba\'nski heavenly equation with a real K\"ahler potential. The
generalized `metric' to be defined with respect to the complex K\"ahler 
potential is not real, and its only use is to make the integrability manifest,
whereas the true metric given by the real part of the generalized K\"ahler 
potential is no longer self-dual.

It should be noticed that the solutions to the gravitational equations of 
motion (24) are all stationary with respect to the {\it Born-Infeld}-type 
effective action
$$ S= \int d^{2+2}z\,\sqrt{-\det\left(g_{i\bar{j}}+\fracmm{\k\sqrt{2}}{g}
F_{i\bar{j}}\right)}~.\eqno(27)$$
The action $S$ is, however, {\it not} the standard Born-Infeld action~\cite{bi}
since the determinant in eq.~(27) is two-dimensional, not four-dimensional.

It is worth mentioning that the infinite hierarchy of conservation laws and the
infinite number of symmetries \cite{bpl2} exist as the consequences of 
Penrose's twistor construction when it is formally applied to our `almost 
self-dual' gravity with a complex K\"ahler potential. The underlying symmetry 
is known to be a loop group $S^1\to SDiff(2)$ of the {\it area-preserving} 
(holomorphic) diffeomorphisms (of a 2-plane), which can be considered as a 
`large N limit' $(W_{\infty})$ of the $W_{\rm N}$ symmetries in 
two-dimensional conformal field theory~\cite{bakas}. The area-preserving
 holomorphic diffeomorphisms, 
$$ \pa_i\bar{\pa}_{\bar{j}}\ck(z,\bar{z})\to \pa_i\x^k(z)
\pa_k\bar{\pa}_{\bar{k}}\ck(\x,\bar{\x})\bar{\pa}_{\bar{j}}\bar{\x}^{\bar{k}}
(\bar{z})~, \eqno(28)$$
leave eq.~(27) to be invariant, since $ \abs{\det(\pa_i\x^k)}=1$
by their definition. It is, therefore, natural to interpret eq.~(27) as a 
(part of) particular {\it F-brane} action whose world-volume is 
(2+2)-dimensional. 

The apparent drawback of the F-brane action (27) is the lack of `Lorentz'
covariance in the four-dimensional world-volume. It can, however, be corrected
by supersymmetrizing the action (27). Like that in superstring theory, there 
may be two different ways of supersymmetrization, either in the world-volume
({\it a l\'a} Neveu-Schwarz-Ramond), or in the F-brane target space ({\it 
a l\'a} Green-Schwarz). Leaving aside the second possibility here (see, 
however, ref.~\cite{brane} for some proposals), let's consider the first one 
which amounts to the (N-extended) supersymmetrization of self-dual gravity. 
The latter can be constructed by promoting an $SL(2,R)'$ factor of the 
(2+2)-dimensional `Lorentz' group $SO(2,2)\cong SL(2,R)\otimes SL(2,R)'$ to 
the supergroup $OSp(N|2)$ to be locally realised, while keeping the rigid
symmetry $SL(2,R)$ intact. The on-shell superspace constraints defining the 
N-extended {\it self-dual supergravity} (SDSG) are known, and they can be 
solved in a light-cone gauge, in terms of a self-dual N=8 superfield 
potential $V_{=,=}$ of helicity (-2)~\cite{sie}. The non-covariant but
manifestly N-supersymmetric SDSG action in terms of the potential $V_{=,=}$ 
takes the form of the Pleba\'nski action in the light-cone N-extended 
self-dual superspace~\cite{sie}. The covariant {\it and} manifestly 
supersymmetric SDSG action can {\it only} be constructed in a harmonic 
superspace when the supersymmetry is the {\it maximal} one, i.e. when N=8 
\cite{prep}. The N=8 harmonic superspace has extra bosonic coordinates that 
parametrize a coset $SL(2,R)/GL(1)$~\cite{prep}. The $SO(8)$ symmetry rotating
32 supercharges is automatically gauged in the N=8 SDSG. The latter has, 
therefore, the bosonic symmetry
$$ \left[ SO(1,2)\otimes SO(8)\right]_{\rm local}\otimes\left[ 
SL(2,R)\right]_{\rm global}~.\eqno(29)$$
This symmetry should have a natural explanation as the broken Lorentz
symmetry of yet another supergravity theory in higher dimensions. A natural
candidate for the higher-dimensional supersymmetric generalization of the 
(2+2)-dimensional `Lorentz' group is the supergroup $OSp(32|1)$ whose basic 
anticommutation relation in 2+10 dimensions reads
$$ \{ Q_{\a},Q_{\b}\}=\g^{\m\n}_{\a\b}M_{\m\n}+\g_{\a\b}^{\m_1\cdots\m_6}
Z^+_{\m_1\cdots\m_6}~.\eqno(30)$$
In eq.~(30), the 32 supercharges $Q_{\a}$ comprise a Majorana-Weyl spinor,
the twelve-dimensional gamma matrices are all chirally projected, the 66 
generators $M_{\m\n}$ represent a two-form, and the 462 generators 
$Z^+_{\m_1\cdots\m_6}$ is a self-dual six-form, all in 2+10 dimensions. 
The superalgebra $OSp(32|1)$ can be interpreted as either (i) the self-dual
(1,0) supersymmetric or `heterotic' Lorentz superalgebra in 2+10 dimensions,
or (ii) de Sitter supersymmetry algebra in 1+10 dimensions, or (iii) the 
conformal supersymmetry algebra in 1+9 dimensions~\cite{brane}. 

The most striking feature of the superalgebra (30) is that it is {\it not} of 
the super-Poincar\'e type, since it does not contain the translation operators
on the right-hand-side of eq.~(30) in twelve dimensions. Nevertheless, they
do appear after the Wigner-In\"on\"u contraction down to eleven 
dimensions.~\footnote{The 66 generators $M_{\m\n}$ can be decomposed in eleven
dimensions to 55 `Lorentz' generators and 11 translation \newline ${~~~~~}$ 
generators.} Eq.~(29) also suggests that the hypothetical heterotic (1,0) 
supergravity may not be fully Lorentz-covarint in 2+10 dimensions. The 
$SL(2,R)$ rigid symmetry in eq.~(29) should be interpreted as a {\it duality} 
symmetry. Its discrete subgroup $SL(2,Z)$ is going to survive as the 
strong-weak coupling duality in the F-theory whose low-energy limit should be 
our `heterotic' twelve-dimensional supergravity. 
\vglue.2in

\noindent {\bf Acknowledgements}\\
I would like to thank the Organizers of the Conferences in Brisbane and Tomsk
for their kind invitations and a nice atmosphere at the meetings. I am also
grateful to Louise Dolan, Jim Gates, Murat G\"unaydin, Evgeny Ivanov, Hitoshi 
Nishino and Paul Tod for useful discussions. 
\vglue.2in


\noindent
\begin{thebibliography}{99}
\bibitem{ov} H. Ooguri and C. Vafa, Nucl. Phys. B361 (1991) 469.
\bibitem{mar} N. Marcus,  Nucl. Phys. B387 (1992) 263.
\bibitem{ovh}  H. Ooguri and C. Vafa, Nucl. Phys. B367 (1991) 83.
\bibitem{svk} S. V. Ketov, Phys. Lett. 395B (1997) 48.
\bibitem{klm} D. Kutasov and E. Martinec, Nucl. Phys.  B477 (1996) 652.
\bibitem{hann} J. Bischoff, O. Lechtenfeld and S. V. Ketov, Nucl. Phys. B438
 (1995) 373.
\bibitem{bv} N. Berkovits and C. Vafa, Nucl. Phys.  B433 (1995) 123.
\bibitem{dns} S. Donaldson, Proc. London Math. Soc. 50 (1985) 1;\\
V. P. Nair and J. Schiff, Nucl. Phys. B371 (1992) 329. 
\bibitem{yang} C. N. Yang, Phys. Rev. Lett.  38 (1977) 1377; Phys. Lett.
 84B (1979) 411.
\bibitem{lez} A. N. Leznov, Theor. Math. Phys.  73 (1988) 1233.
\bibitem{parkes} A. Parkes, Phys. Lett.  286B (1992) 265.
\bibitem{slecht} O. Lechtenfeld and W. Siegel, Phys. Lett. 405B (1997) 49.
\bibitem{pen} R. Penrose, Gen. Rel. Grav.  7 (1976) 31.
\bibitem{fla} E. J. Flaherty, Gen. Rel. Grav.  9 (1978) 961.
\bibitem{bpl} C. P. Boyer and J. F. Pleba\'nski, Phys. Lett. 106A (1984) 125.
\bibitem{bi} M. Born and L. Infeld, Proc. Roy. Soc.  A144 (1934) 425.
\bibitem{bpl2} C. P. Boyer and J. F. Pleba\'nski, J. Math. Phys. 26 (1985) 229.
\bibitem{bakas} I. Bakas, Phys. Lett.  228B (1989) 57; Commun. Math. Phys.
 134 (1990) 487.
\bibitem{brane} S. V. Ketov, Mod. Phys. Lett. A11 (1996) 2369.
\bibitem{sie} W. Siegel, Phys. Rev. D46 (1992) 3225; ibid. D47 (1993) 2504; 
2512.
\bibitem{prep} S. Karnas and S. V. Ketov, in preparation.
\end{thebibliography}
\end{document}